\documentclass[journal,]{IEEEtran}

\ifCLASSINFOpdf
\else
   \usepackage[dvips]{graphicx}
\fi
\usepackage{url}

\usepackage{graphicx}

\usepackage{CJKutf8}

\usepackage[T1]{fontenc}
\usepackage{amsmath}
\usepackage[cmintegrals]{newtxmath}
\usepackage{bm}

\usepackage{booktabs}
\usepackage{makecell}
\usepackage{xspace} 
\usepackage{listings}
\usepackage{xcolor}
\usepackage[caption=false,font=footnotesize]{subfig}
\usepackage{multirow}

\usepackage{float}
\usepackage{graphicx}

\usepackage{flushend}

\newcommand{\IRPSs}{Interleaved Radar Pulse Streams\xspace}
\newcommand{\irps}{interleaved radar pulse streams\xspace}

\newcommand{\sprs}{single radar pulse stream\xspace}

\newcommand{\WVE}{Wide-Value-Embeddings\xspace}
\newcommand{\wve}{WVEmbs\xspace}

\newcommand{\wvd}{wide-value dimension\xspace}
\newcommand{\SLE}{Learn-able Embeddings\xspace}
\newcommand{\sle}{LEmbs\xspace}
\newcommand{\vd}{Value Dimension\xspace} 

\begin{document}
\begin{CJK}{UTF8}{gbsn}

\title{\wve with its Masking: A Method For Radar Signal Sorting}
\author{
Xianan Hu, 
Fu Li, \IEEEmembership{Member, IEEE},
Kairui Niu, 
Peihan Qi, \IEEEmembership{Member, IEEE},
and Zhiyong Liang,
\thanks{
This work was supported in part by 
the National Basic Scientific Research of China under Grants JCKY2023110C099, and the National Natural Science Foundation of China under Grants 62171334.
}
\thanks{
Xianan Hu and Fu Li are with the School of Artificial Intelligence, Xidian University, Xi'an 710126, China 
(e-mail: 
huxianan@stu.xidian.edu.cn, 
fuli@mail.xidian.edu.cn).
}
\thanks{
Kairui Niu, Peihan Qi, and Zhiyong Liang are with the School of Telecommunications Engineering, Xidian University, Xi'an 710126, China (e-mail: nkairui@126.com, 
phqi@xidian.edu.cn,
23011210457@stu.xidian.edu.cn)
}
}

\maketitle

\begin{abstract}

Radar signal sorting (RSS) is crucial in electronic reconnaissance, yet many current deep learning-based RSS methods are often a simple grafting of general tasks without considering the unique characteristics of radar signals. 
Our study proposes a novel embedding method, \WVE (\wve), for processing Pulse Descriptor Words (PDWs) as normalized inputs to neural networks. This method adapts to the distribution of interleaved radar signals, ranking original signal features from trivial to useful and stabilizing the learning process. 
To address the imbalance in radar signal interleaving, we introduce a value dimension masking method on \wve, which automatically and efficiently generates challenging samples, and constructs interleaving scenarios, thereby compelling the model to learn robust features.
Experimental results demonstrate that our method is an efficient end-to-end approach, achieving high-granularity, sample-level pulse sorting for high-density interleaved radar pulse sequences in complex and non-ideal environments.

\end{abstract}


\IEEEpeerreviewmaketitle




\section{Introduction}


\IEEEPARstart{R}{adar} signal sorting (RSS), also known as deinterleaving, is a crucial research topic in electronic reconnaissance. Existing RSS algorithms have limited practical application, often failing to address non-ideal conditions such as high density and significant pulse interleave. 
These methods often assume fully detected pulse signals with a high signal-to-noise ratio, which is rarely the case in real-world scenarios. Electromagnetic interference typically results in a low signal-to-noise ratio (SNR), while a congested electromagnetic environment leads to high pulse interleave and significant pulse loss. These non-ideal conditions cause existing models to perform poorly compared to their theoretical capabilities, greatly limiting their practical application.
Therefore, there is an urgent need for a method that can quickly and accurately extract and identify target signals of interest from complex, aliased signals.

In recent years, deep learning (DL) models have shown superior performance on common time series tasks \cite{liu2024itransformer,wu2023timesnet,wang2024timemixerdecomposablemultiscalemixing,zhou2021informerefficienttransformerlong,wu2022autoformerdecompositiontransformersautocorrelation}. 
However, DL-based RSS are highly specific tasks and remain largely unexplored.
Previous work has explored various methods for radar signal processing. 
For instance, GRU-based RNNs have been used to deinterleave radar pulse streams by addressing sequence patterns with a one-hot vector representation for PDWs to digitize the input, converting deinterleaving into a multi-step binary classification problem through iterative denoising\cite{Classification-Denoising-Deinterleaving}. 
MaskRCNN \cite{10148581} and Residual GNN \cite{10155167} have also been applied to solve RSS tasks. 
Some work \cite{Multi-Task, 9500447} approach radar signal classification and representation as a multi-task learning problem, leveraging the relationships between tasks.
Furthermore, a multistep segmentation method has been found to improve the accuracy of deinterleaving \cite{Multi-Stage}.



\begin{figure*}[t]
    \centering
    \begin{minipage}{0.7\linewidth}
        \includegraphics[width=\linewidth]{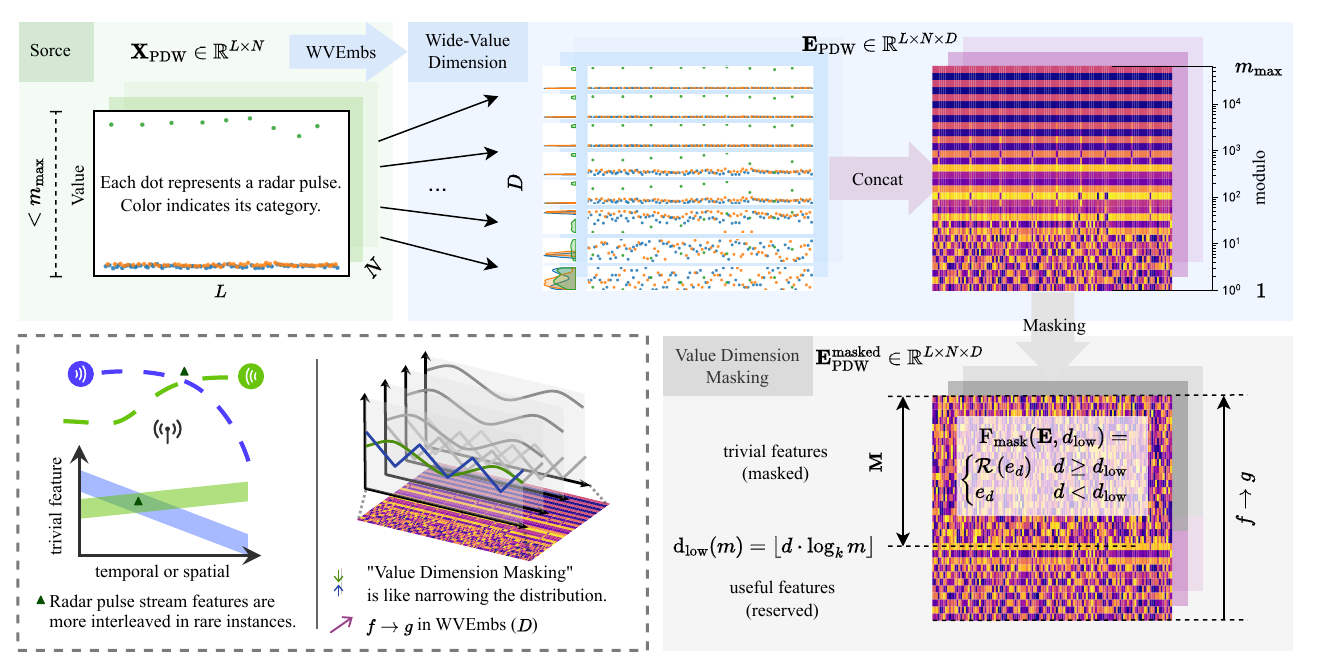}
        \caption{
        The main idea of this paper: 
        1. To address the low overlap probability of trivial features in \irps and the challenges in predicting and preprocessing their distributions (e.g., DOA features overlapped when signal sources are nearby, as shown in the lower left panel). 
        \wve processes and sorts the features of the original signal from trivial to useful in the ``\wvd'', which with each dimension normalized to a 0-1 range, converted from the widely-scaled and sparse original signal, suitable for neural network processing (as shown above, note the visibility of the green radar pulses in the \wve). 
        The masking process helps the classifier focus on learning useful features, mitigating the risk of training crashes due to the predominance of trivial features (bottom right).
        }
        \label{fig:mainfig}
    \end{minipage}
    \quad
    \begin{minipage}{0.25\linewidth}
        \includegraphics[width=\linewidth]{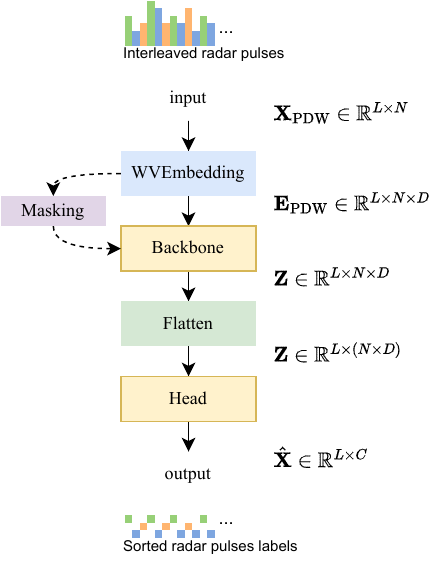}
        \caption{Pipeline of our model used for RSS. The backbone is stacked ModernTCN blocks.
        Masks are applied with a certain probability $P$ only during model training. Experimental results indicate that $P$ has minimal impact on the model's performance.
        }
        \label{fig:pipeline}
    \end{minipage}
    
\end{figure*}

In this paper, as shown in Figure \ref{fig:mainfig}, we introduced the embedding method, \WVE(\wve), and the hard sample mining method, Masking. These techniques can retain more original signal information, construct challenging samples automatically and efficiently, and compel the model to learn robust features.
Due to issues such as the difficulty in training TF-based methods because of the lack of priors like ``time-translation invariance" and the ease of losing temporal information \cite{Zeng_Chen_Zhang_Xu_2023}, we use the CNN-based ModernTCN\cite{donghao2024moderntcn} as the backbone classifier,
Our proposed method can solve the RSS problem in a one-step, end-to-end manner, without requiring a separate clustering step as in \cite{9292936} and \cite{9667203}.
Simulation results show that our method achieves an RSS accuracy of 88.680\% on our dataset, improving performance by about 5dB in three different scenarios. Ablation experiments further demonstrate that our method outperforms existing general models when they are applied to RSS.

The rest of this paper is organized as follows. 
Section \ref{sec:method} illustrates our method according to the characteristic of \irps.
Simulations are carried out in section \ref{sec:exp} to demonstrate the performances of our method. 
Section \ref{sec:conclusion} concludes the whole paper.

\section{Proposed Method}
\label{sec:method}


In radar systems community, related research on radar signals relies on conventional radar parameters for identification, with the most common method being the use of Pulse Descriptor Words  
(PDWs, 
${\mathbf X}_{\text{PDW}}=\{\text{PDW}_1,...,\text{PDW}_L\}$
$\text{PDW}_i=\{{\text{TOA}}_i, \text{RF}_i, \text{PW}_i, \text{PA}_i, \text{DOA}_i\}$)\footnote{
TOA ($\text{TOA}_i$), Time Of Arrival of the pulse,
RF ($\text{RF}_i$), Radio Frequency,
PW ($\text{PW}_i$), Pulse Width,
PA ($\text{PA}_i$), Pulse Amplitude,
DOA ($\text{DOA}_i$), Direction of Arrival,
Different datasets may lack or contain certain parameters. Note that many studies\cite{SDIF,cumulative_difference_histogram} tend to calculate the Pulse Repetition Interval (PRI, $\text{PRI}_i$) as a feature of PDWs using differential TOA. However, due to interleaving, the original PRI feature cannot be directly obtained, so we do not consider PRI.
}. 
In addition, we have the corresponding number of \sprs classes present (${\mathbf X}_{\text{c}}$). 
The dataset we use assumes that each
$\text{PDW}_i$ belongs to a single signal.
Our goal for RSS is to deinterleave and classify PDW streams in a single step simultaneously, 
which is similar to the ``token classification'' task in the NLP field.

Unlike \sprs, PDWs from \irps present unique challenges: Non-interleaved signals are easier to preprocess, but interleaved signal streams are \sprs arbitrarily combined, varying widely with different interleavings. 
Normalizing PDWs' distribution using conventional methods can disrupt the distribution of individual signals, compromising feature learnability.
Furthermore, the types of interleaved patterns increase exponentially with the number of categories of the original radar signal. 
In most cases, interleaving can be achieved through direct clustering; 
however, highly interleaved scenarios as hard examples, are sparse in the dataset and require significant domain expertise to construct manually. 
Preventing the vast number of easy samples from overwhelming the classifier during training remains a challenge.
To sum up, our method is designed to address the following challenges in RSS:


\begin{itemize}
    \item Processing PDWs effectively across a very wide range of scales to facilitate learning by NNs without heavy reliance on prior assumptions.
    \item Reevaluating normalization techniques due to the unpredictable distribution of \irps.
    \item Mining hard samples without compromising the accuracy of NNs on simpler cases.
\end{itemize}

\subsection{\wve Generation for \IRPSs}

\subsubsection{Preliminaries}

Preprocessing input data in NNs is crucial due to distributional differences between training and test data.
Some recent work, such as Series Stationarization and De-stationary Attention\cite{liu2023nonstationarytransformersexploringstationarity} and using normalization method with learnable affine transformation \cite{RevIN} to adapt to the input distribution and preserve non-stationary information. But they still treat the input signal as a normalized whole.
The concept of embeddings was initially used to compress feature dimensions and stabilize the learning process. However, we observe that using \SLE (\sle) can cause severe oscillations during training and hinder convergence, indicating a steep loss surface and poor model robustness (ref. Section \ref{sec:ablation}).
Some related work involves sampling the original signal into word embeddings\cite{Classification-Denoising-Deinterleaving}. This digitizing approach causes ``resolution problems\cite{SSD}'' which may introduce sampling loss and disrupt the continuous nature of the original signal.
We need a specialized embedding method to achieve both normalization and embedding while keeping important non-stationary information.

\subsubsection{Designing}




We build our embedding (\wve) through the following methods:
Assume that the $N$ dimensional variable ${\mathbf X}_{\text{PDW}} \in \mathbb {R}^{L \times N}$ of length $L$ is the input signal. $D$ is the desired dimension of the embedding, and $E_{i} \in \mathbb{R}^{D}$ is the embedding. 
Then we can set the step size $\delta  \in \mathbb{N}^+$ and the scaling factor $k \in \mathbb{N}^+$ to control the upper bound of the maximum modulus $M = k^{D/\delta }$ or we can determine $\delta  =  D \log_{M}{k}$. 
We need to apply an affine transformation to the input data during preprocessing to ensure that the data information is reflected in the residue system with moduli ranging from $1$ to $m_{\mathrm{max}}, (m_{\mathrm{max}} = M / k^{1}, 1 \ll m_{\mathrm{max}})$.
Since $M \propto k^D$, we can easily choose $d,k$ to satisfy this condition.
Thus, we obtain \wve using the following formula: ${\mathbf E}_{\text{PDW}} \in \mathbb {R}^{L \times N \times D}$:

\begin{gather}
    {\mathbf E}_{\text{PDW}}^{l,n,\delta  i+j} =
    f_{\omega =1}(\frac{{{\mathbf X}^{l,n}_{\text{PDW}}}}{M^{\delta  i/D}} + \frac{j}{d})  \\
    \delta \mid D,\; i \in \{0,1,...,D/\delta  -1\} ,\; j\in \{1,2,...,\delta \} \notag
\end{gather}

where $di+j\in \{1,2,...,D\}$  is a dimension of the embedding and ${\mathbf X}^{l,n}_{\text{PDW}}$ is the value of the input token.
$f_{\omega =1}(\cdot)$ is the linear periodic function with a period of 1 that is used, we use\footnote{
We chose this function, rather than the trigonometric functions used in positional encoding, as a source of inspiration\cite{Attention}, Because we assume that in the convolution operation, it can make the model easily learn relative attention because for any fixed value $x$ offset by $k$, we want ${\mathbf E}(x)$ to be approximately the same as ${\mathbf E}(x+k)$. If the backbone uses a structure similar to Attention, then trigonometric functions should be used.
}:

\begin{equation}
    f_{\omega =1}(x) \triangleq x \bmod 1 \cdot 2-1
\end{equation}

In other words, \wve is equivalent to a family of periodic functions with period $k$ from $1$ to $M$. $k \in \mathbb{N}^+$ required to prevent information leakage in subsequent masking.
We consider that
\wve implicitly performs multi-periodic decomposition of the time dimension. Compared to widely used timestamp preprocessing methods, such as hierarchical timestamps \cite{zhou2021informerefficienttransformerlong}, \wve eliminates the need for tedious manual preprocessing based on a priori knowledge and avoids complex FFT calculations \cite{wu2023timesnet}. And it inherently includes seasonal trend decomposition \cite{wu2022autoformerdecompositiontransformersautocorrelation}.
\wve generates one dimension from a single value, which is a variable independent embedding, suitable for subsequent backbone operations such as separable spatio-temporal convolutions.

\subsection{Hard Sample Mining Based on Value Dimension Masking}



\subsubsection{Preliminaries}

Masking or disrupting the input \cite{jing2022masked, he2021maskedautoencodersscalablevision, devlin2019bertpretrainingdeepbidirectional} remains one of the simplest and most effective methods to prevent NNs from converging to the local optima by classifying simple features, especially when hard examples are sparse in the dataset.
It can increase the proportion of difficult samples in the training data, thereby forcing the backbone NN to classify signals based on deeper sequential features rather than relying on superficial, trivial statistical characteristics.


\subsubsection{Designing}

We refer to the dimension extended by \wve as the ``\vd'', which we apply masking $\mathbf{M}$ to the \vd. We the filling values for the mask area as $\mathbf{z}$. 
For a useful feature $f$ and a trivial feature $g$, 
we require $\mathbf{M}$ and $\mathbf{z}$  to satisfy: 
        $||f(\mathbf{M}*\mathbf{x} + (1-\mathbf{M}) * \mathbf{z}) - f(\mathbf{M}'*\mathbf{x} + (1-\mathbf{M}') * \mathbf{z}')||^2 \approx ||f(\mathbf{x}_1) - f(\mathbf{x}_2)||^2$ and $||g(\mathbf{M}*\mathbf{x} + (1-\mathbf{M}) * \mathbf{z}) - g(\mathbf{M}'*\mathbf{x} + (1-\mathbf{M}') * \mathbf{z}')||^2 \gg ||g(\mathbf{x}_1) - g(\mathbf{x}_2)||^2$.
\footnote{
Useful feature, such as time series features from a \sprs, and trivial feature, such as simple statistical features, are described in detail in Figure \ref{fig:mainfig}.
This masking scheme is inspired by work on representation learning with Siamese networks \cite{jing2022masked}, but we incorporate the constructed hard samples directly into the training set. Its application to unsupervised learning is also a for future research.
}
Due to the proper design of our \wve, a higher embedding dimension $d$ means that features change from useful to trivial, $f \to g$, As shown on the right side of Figure \ref{fig:mainfig}, The semantics implied by lower dimensions are more complex and can even tend toward random values, whereas higher-dimensional features are more distinct and even can be directly clustered.

Given a modulus $m$, find the lowest masking dimensional function ${\mathrm d}_{\mathrm {low}}(\cdot)$ is:

\begin{align} 
{\mathrm d}_{\mathrm {low}}(m) = \left \lfloor \delta \cdot \log_{k} m \right \rfloor
\end{align}

The mask function $F_\mathrm{mask}(\cdot, d_{\mathrm{low}})$ takes ${\mathbf E}_{\text{PDW}} \in \mathbb {R}^{L \times N \times D}$, as input and outputs ${\mathbf E}_{\text{PDW}}^{\mathrm{masked}} \in \mathbb {R}^{L \times N \times D}$ which is a mask on the time dimension.

\begin{equation}
    \mathrm{F}_\mathrm{mask}({\mathbf E}, d_{\mathrm{low}})=\begin{cases}
	\mathcal{R} \left( e_d \right)& d\ge d_{\mathrm{low}}\\
	e_d&  d<d_{\mathrm{low}}\\
    \end{cases}
    \label{equ:mask}
\end{equation}

Here, $E_{}=[ e_{1}^{l,n},e_{2}^{l,n},...,e_{D}^{l,n} ]$, $d_{\mathrm{low}} = {\mathrm d}_{\mathrm {low}}(m), m \sim P$, and $P$ represents the distribution of the maximum and minimum differences of the original single-class radar $\text{PDW}_{n}$ features. 
$\mathcal{R}(\mathbf{e}) \sim U(0, 1)^{\mathrm{shape}(e)} $
, as the \texttt{rand\_like} function is provided in common DL libraries such as NumPy \cite{harris2020array} or PyTorch \cite{Ansel_PyTorch_2_Faster_2024}.









\section{Experiments}
\label{sec:exp}

\subsection{Datasets}



Existing widely used datasets are either oriented towards simple radar signal classification tasks \cite{8267032,9500447,Multi-Task}, do not account for the interleaving of radar signals in the time domain, or only consider raw IQ signals \cite{Multi-Stage}.
To demonstrate the strong deinterleaving capability of our method, we use a dataset
\footnote{From the 
2nd Electromagnetic Big Data Super Contest.
} containing twelve types of radar signals, as shown in Table \ref{tab:data}. This dataset also provides three scenarios as testing sets to simulate \irps signals in real-world environments.

\begin{table}[h]
        \centering
        \caption{Parameter Settings For Each Radar Emitter}
        \label{tab:data}
        \small
        \setlength{\tabcolsep}{4pt}
        \scalebox{0.8}{
        \begin{tabular}{ccccccc}
        \toprule
            Radar & DOA/° & PW/$\mu$s & RF/MHz & PRI/$\mu$s & PA/dBm & Pulse No. \\ 
        \midrule
            1  & 54$\sim$61 & 8$\sim$9   & 9591$\sim$9619 & 37$\sim$147  & -102$\sim$0 & 129,720 \\
            2  & 39$\sim$47 & 8$\sim$10  & 9610$\sim$9620 & 37$\sim$148  & -115$\sim$0 & 120,110 \\
            3  & 38$\sim$45 & 9$\sim$10  & 9606$\sim$9612 & 37$\sim$63   & -135$\sim$0 & 193,320 \\
            4  & 44$\sim$51 & 8$\sim$10  & 9636$\sim$9643 & 37$\sim$121  & -103$\sim$0 & 133,054 \\
            5  & 45$\sim$52 & 9$\sim$11  & 9592$\sim$9625 & 37$\sim$113  & -117$\sim$0 & 137,488 \\
            6  & 43$\sim$51 & 8$\sim$10  & 9579$\sim$9596 & 36$\sim$111  & -114$\sim$0 & 143,848 \\
            7  & 57$\sim$63 & 9$\sim$11  & 9557$\sim$9564 & 37$\sim$98   & -109$\sim$0 & 155,580 \\
            8  & 58$\sim$65 & 9$\sim$10  & 9566$\sim$9616 & 7$\sim$41    & -132$\sim$0 & 222,870 \\
            9  & 57$\sim$64 & 9$\sim$10  & 9558$\sim$9613 & 16$\sim$52   & -120$\sim$0 & 150,702 \\
            10 & 51$\sim$58 & 9$\sim$10  & 9575$\sim$9580 & 24$\sim$44   & -112$\sim$0 & 120,392 \\
            11 & 38$\sim$45 & 1$\sim$2   & 9579$\sim$9653 & 7$\sim$39    & -132$\sim$0 & 157,934 \\
            12 & 49$\sim$55 & 33$\sim$38 & 2825$\sim$2841 & 398$\sim$508 & -91$\sim$0  & 22,112  \\
        \midrule
            I & 43$\sim$67 & 1$\sim$8   & 9725$\sim$9743 & 0$\sim$118   & -128$\sim$0 & 999,038 \\
            II & 50$\sim$286& 1$\sim$9   & 9082$\sim$9118 & 0$\sim$143   & -129$\sim$0 & 1,762,620 \\
            III & 34$\sim$66 & 1$\sim$39  & 2807$\sim$9617 & 0$\sim$116   & -149$\sim$0 & 2,579,053 \\
        \bottomrule
        \end{tabular}
        }
\end{table}

\subsection{Model Structure}

We input the preprocessed \irps as ${\mathbf X}_{\text{PDW}} \in \mathbb {R}^{L \times N}$, with the corresponding labels being ${\mathbf X}_{\text{class}} \in \mathbb {R}^{L \times C}$, Through our \wve, the \irps will be embedded into D-dimensional embedding vectors:

  
\begin{equation}
    \mathbf{X}_\text{emb} = \text{WVEmbedding}(\mathbf{X}_\text{in})
\end{equation}

  
After embedding, We use stacked ModernTCN blocks  \cite{donghao2024moderntcn} with residual connections as the $\text{Backbone}(\cdot)$. \wve is fed into the backbone to capture both the cross-time and cross-variable dependency and learn the informative representation ${\mathbf Z} \in \mathbb {R}^{L \times N \times D}$ :

  
\begin{equation}
    \mathbf{Z} = \text{Backbone}(\mathbf{X}_\text{emb})
\end{equation}

  
The classification head adheres to the general design of token classification and utilizes a linear probing layer only on the flattened ${\mathbf Z} \in \mathbb {R}^{L \times (N \times D)}$, Then a projection layer with SoftMax activation is to map the final representation to the final classification result ̂$\hat{\mathbf X} \in \mathbb {R}^{L \times C}$, where $C$ is the number of classes.


\begin{equation}    
    \hat{\mathbf X} = \text{Softmax} \Big (\text{LinearProbing}\big(\text{Flatten}(\mathbf{Z})\big)\Big)
\end{equation}

  

We optimize the network using cross-entropy loss and the AdamW optimizer \cite{loshchilov2019decoupledweightdecayregularization}. The backbone network employs BatchNorm \cite{BatchNormalization,santurkar2019doesbatchnormalizationhelp}, which facilitates faster and more stable training. Dropout \cite{Dropout,li2018understandingdisharmonydropoutbatch} is applied only in the classification head to regularize the model. Figure \ref{fig:pipeline} illustrates the overall network architecture.


\subsection{Training Details}


We trained the models using a single NVIDIA GeForce RTX 3090 GPU. 
Our data augmentation process involves random interceptions, drops, and affine transformations applied to 12 types of signals, recombined into simulated interleaved signals to mimic real-world scenarios. 
This process helps preserve the essential information of signals within the \wve generation without needing normalization. We apply a fixed affine transformation, ${a} \cdot {\mathbf X^n} + {b}, \forall {\mathbf X^n} \in {\mathbf X}_{\text{PDW}}$, where the coefficients ${a}$ and ${b}$ are adjusted based on the signal distribution to maintain the integrity of meaningful information.

\subsection{Model Performance}

\begin{figure}[ht]
    \centering
    \subfloat[Scene I]{\includegraphics[width=0.3\linewidth]{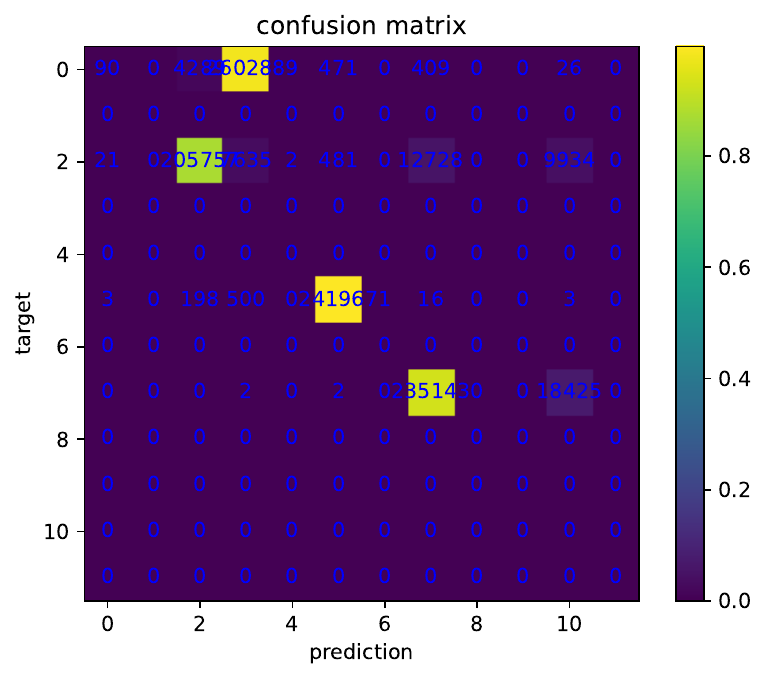}}
    \quad
    \subfloat[Scene II]{\includegraphics[width=0.3\linewidth]{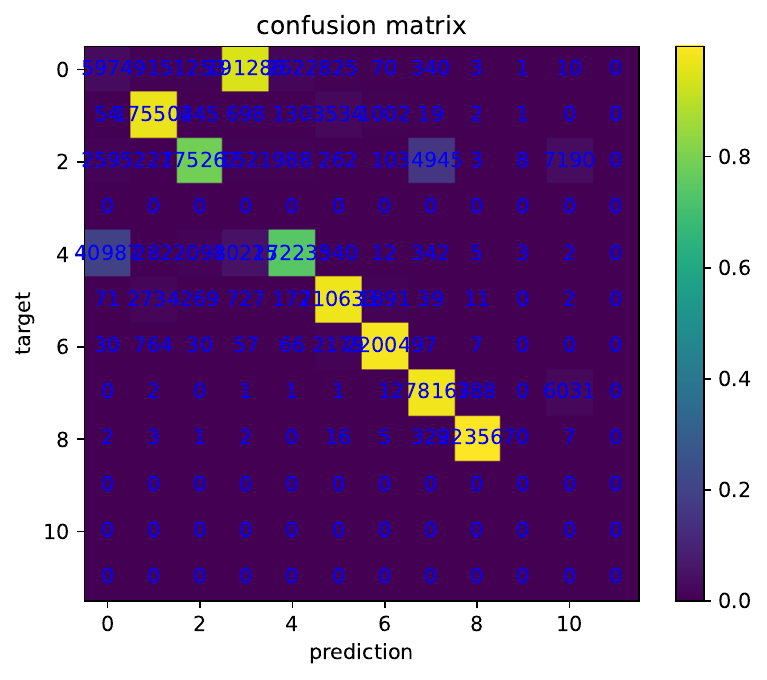}}
    \quad
    \subfloat[Scene III]{\includegraphics[width=0.3\linewidth]{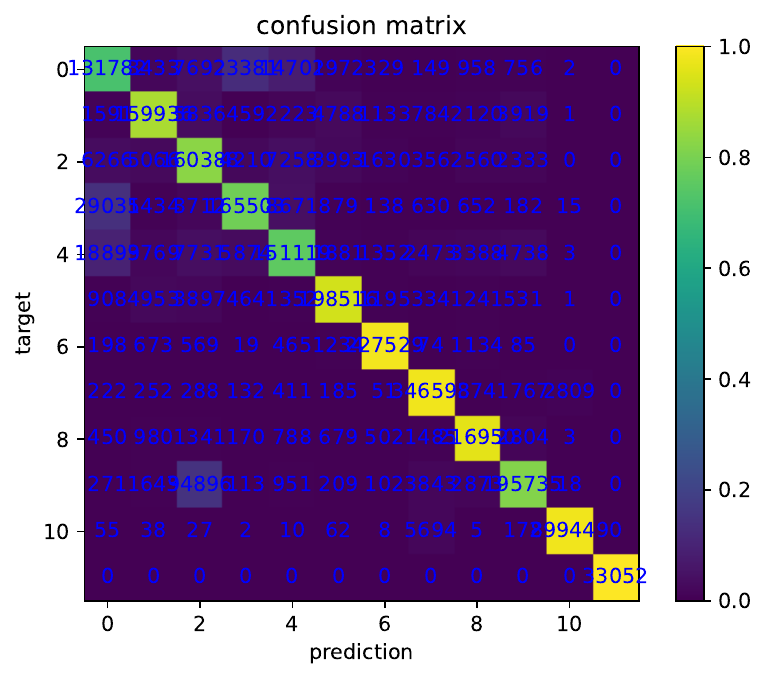}}
    \caption{Confusion matrices for the three test scenes.}
    \label{fig:matrix}
\end{figure}

Consider the potential imbalance of sampling points. we adopt the precision ($\mathrm{Pre}={\frac {\mathrm{TP}}{\mathrm{TP}+\mathrm{FP}}}$), recall ($\mathrm{Rec}={\frac {\mathrm{TP}}{\mathrm{TP}+\mathrm{FN}}}$) and F1-score ($F_{1}={\frac {2}{\mathrm{Rec} ^{-1}+\mathrm{Pre} ^{-1}}}$) to evaluate the performance of classification tasks.
The confusion matrix of our proposed model for the three test scenarios in the dataset is shown in Figure \ref{fig:matrix}. The classification performance is shown in table \ref{tab:ablation}.

\subsection{Ablation Study}
\label{sec:ablation}
\begin{figure}[h]
    \centering
    \includegraphics[width=\linewidth]{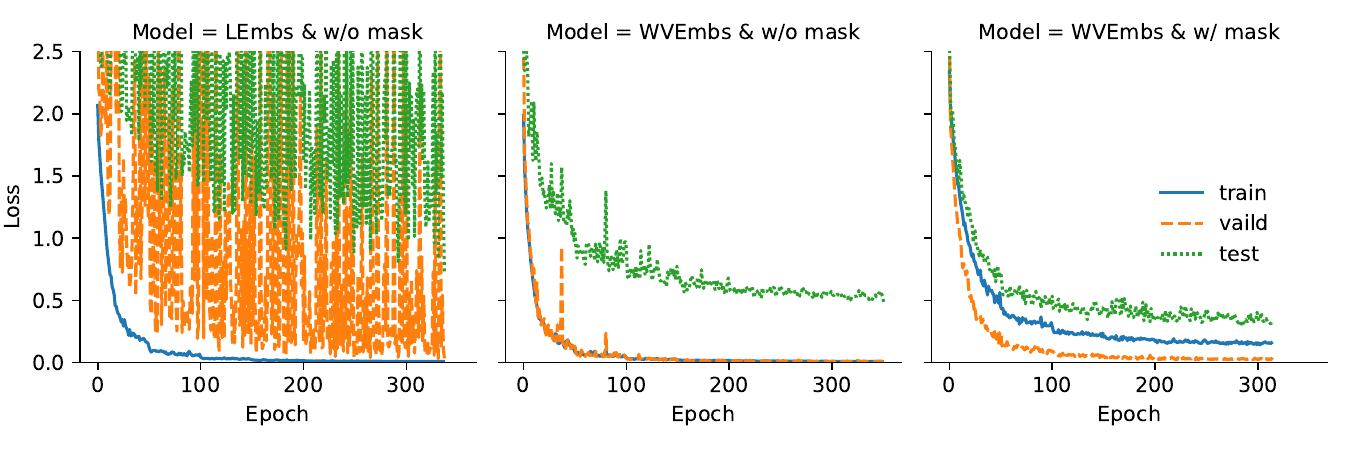} \\
    \includegraphics[width=\linewidth]{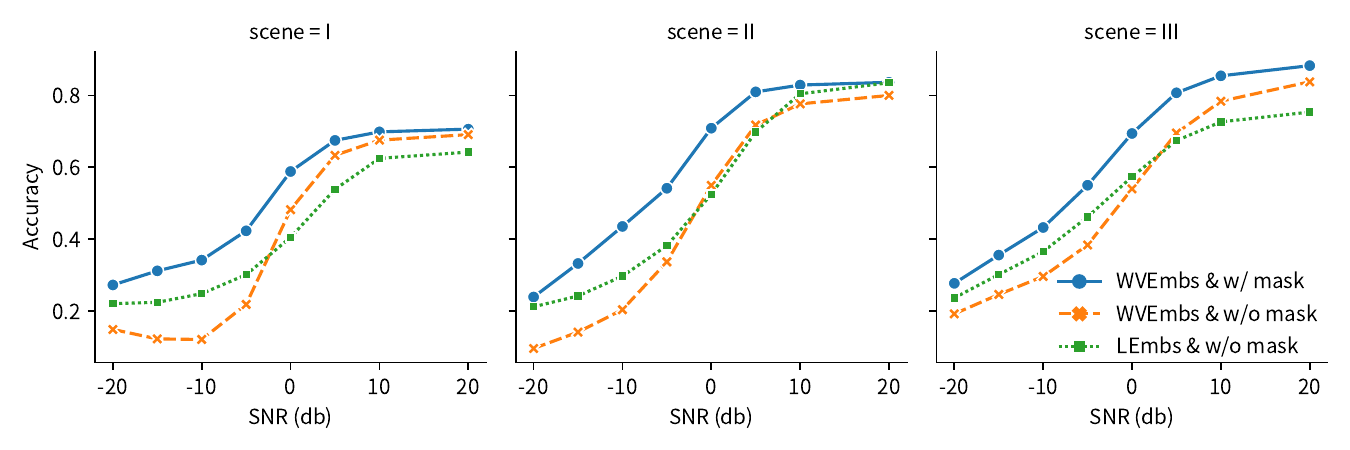}
    \caption{Above: \wve significantly enhances the stability of the training process. \vd masking facilitates loss convergence and improves accuracy.
        Below: Model performance across an SNR range of -20 to 20 dB in three scenarios.
        }
    \label{fig:snr}
\end{figure}

To demonstrate that \wve with its masking method improves model performance compared to the default general \sle, we conducted an ablation study, as shown in Table \ref{tab:ablation}, we additionally normalize the input data window in \sle model.
Note that the model using \wve exhibits slightly higher performance, as the introduction of \wve reduces the model parameters associated with the \sle component.
While \sle theoretically encompass \wve, the fixed-generation \wve often performs better in practice. We speculate that \wve offers more prior knowledge, making it easier for NNs to fit subtle features across wide scales.
As shown in Figure \ref{fig:snr}, our method improves training stability and model robustness significantly.


\begin{table}[ht]
    \centering
    \caption{\wve ablation experiment and comparison of masking methods}
    \label{tab:ablation}
    \scalebox{1}{
    \begin{tabular}{ccccc}
        \toprule
        ~ & ~ & \multirow{2}*{\sle}  & \multicolumn{2}{c}{\wve ~}\\ 
        \cmidrule(l){4-5} 
        ~ & ~ & ~ & w/o mask & w/ mask (ours)\\ 
        \midrule
        \multirow[c]{3}{*}{I} & Pre & 87.116 & 79.495 & \textbf{92.453} \\
         & Rec & 60.633 & 68.242 & \textbf{68.405} \\
         & F1 & 64.348 & 69.389 & \textbf{69.864} \\
        \cmidrule{1-5}
        \multirow[c]{3}{*}{II} & Pre & \textbf{87.373} & 83.775 & 86.547 \\
         & Rec & 81.597 & 78.498 & \textbf{81.790} \\
         & F1 & \textbf{83.760} & 80.062 & 83.488 \\
        \cmidrule{1-5}
        \multirow[c]{3}{*}{III} & Pre & 76.237 & 84.458 & \textbf{88.873} \\
         & Rec & 76.107 & 84.138 & \textbf{88.680} \\
         & F1 & 75.650 & 84.097 & \textbf{88.696} \\
        \bottomrule
    \end{tabular}
    }
\end{table}

\section{Conclusion}
\label{sec:conclusion}


We introduced a method that offers two key advantages:
1. \wve generation efficiently processes and prioritizes PDW features.
2. Hard sample mining through \vd masking automatically generates difficult samples, enhancing robust feature learning.
Results confirm that this method can replace traditional embedding techniques while reducing model parameters, and improving accuracy and robustness. It enables sample-point-level sorting and annotation of dense, interleaved radar pulses in challenging environments. This efficient end-to-end approach avoids the need for data digitization or separate networks for each target type.
Future work should conduct more comparative experiments with other algorithms, and further explore the theoretical foundations of \wve and its application to general time series tasks.






\newpage

\bibliographystyle{IEEEtran}
\bibliography{IEEEabrv,reference}

\end{CJK}
\end{document}